\documentstyle[aps,prb,multicol,epsfig]{revtex}
\begin{document}
\draft
\preprint{}
\wideabs{
\title{Interchain interactions and magnetic properties of Li$_2$CuO$_2$}
\author{Y. Mizuno, T. Tohyama, and S. Maekawa}

\address{Institute for Materials Research, Tohoku University,
        Sendai 980-8577, Japan}

\date{\today}
\maketitle

\begin{abstract}
An effective Hamiltonian is constructed for an insulating cuprate with edge-sharing chains Li$_2$CuO$_2$. The Hamiltonian contains the nearest and next-nearest neighboring intrachain and zigzag-type interchain interactions. The values of the interactions are obtained from the analysis of the magnetic susceptibility, and this system is found to be  described as coupled frustrated chains. We calculate the dynamical spin correlation function $S({\bf q}, \omega)$ by using the exact diagonalization method, and show that the spectra of $S({\bf q}, \omega)$ are characterized by the zigzag-type interchain interactions. The results of the recent inelastic neutron-scattering experiment are discussed in the light of the calculated spectra.

\end{abstract}
}
\narrowtext

One-dimensional cuprates have received much attention as reference systems of high-{\it T$_c$} superconductors with two-dimensional CuO$_2$ planes. Recently, a variety of the compounds with edge-sharing chains, where CuO$_4$ tetragons are coupled by their edges, were synthesized and found to show unique physical properties. Li$_2$CuO$_2$ is one of the typical compounds having such chains.  As shown in Fig.~1(a), the chains run parallel to b-axis and are stacked along a- and c-axes.\cite{Sapina}

An important feature of the edge-sharing chains is that a nearest neighboring (NN) magnetic interaction $J_1$ between Cu spins strongly depends on Cu-O-Cu bond angle $\theta$. In the case that $\theta$ = 90$^\circ$, the superexchange process via O ions, which contributes to antiferromagnetic (AFM) interaction, is suppressed due to the orthogonality of Cu3$d$ and O2$p$ orbitals, and ferromagnetic (FM) contribution caused by, for example, direct exchange mechanism between Cu3$d$ and O2$p$ orbitals, becomes dominant.\cite{Geertsma,Mizuno1} With increasing $\theta$, the AFM superexchange interaction increases, and consequently $J_1$ changes from FM to AFM interaction at a critical angle $\theta_c$. In the previous study,\cite{Mizuno1} $\theta_c$ was estimated to be about 95$^\circ$ from the cluster calculation. For Li$_2$CuO$_2$ with $\theta=94^\circ$, $J_1$ was evaluated to be FM ($<$0) with magnitude of 100~K. In addition to $J_1$, a next-nearest neighboring (NNN) magnetic interaction $J_2$, which comes from Cu-O-O-Cu path, also plays an important role in the magnetic properties. The interaction $J_2$ is AFM ($>$0), and its magnitude is known to be comparable to $|J_1|$. Therefore, an appropriate model describing the edge-sharing chain is a spin 1/2 Heisenberg model with NN and NNN interactions (a $J_1$-$J_2$ model). The ground state of the $J_1$-$J_2$ model has been extensively studied:\cite{J1J2} For $J_2/|J_1|<1/4$, it is a FM state, while for $J_2/|J_1|>1/4$, it is a frustrated state with incommensurate spin correlation.

In Li$_2$CuO$_2$, AFM long-range order occurs at $T_{N}$=9~K, and the magnetic structure below $T_{N}$ is FM along a- and b-axes and AFM along c-axis.\cite{Sapina} The recent inelastic neutron-scattering experiment showed the existance of interchain (IC) interactions which bring about the ordering.\cite{Boehm} The analysis of the dispersions along a- and c-axes in the linear spin-wave theory revealed that the IC interaction between NN Cu spins is of the order of 10~K. The band calculation also showed the existence of large effective hoppings between neighboring chains, and thus the superexchange interactions.\cite{Weht} Following these facts, IC interaction plays an important role in the magnetic properties of Li$_2$CuO$_2$. The importance of IC interactions have also been pointed out in other edge-sharing cuprates, CuGeO$_3$ (Ref.~7) and Sr$_{14}$Cu$_{24}$O$_{41}$ (Refs. 8 and 9).

In this paper, paying attention to the IC interaction, we examine the magnetic properties of Li$_2$CuO$_2$ such as magnetic susceptibility and magnetic excitation by applying the exact diagonalization method on finite size clusters. We construct a minimal model, taking into account the IC interaction to the $J_1$-$J_2$ model. A set of parameter values is obtained from the analysis of the temperature dependence of the magnetic susceptibility $\chi(T)$. We find that the compound is described as a system with frustrated chains coupled by zigzag-type IC interactions. In order to examine the magnetic excitation, we calculate the dynamical spin correlation function $S({\bf q},\omega)$ in the chain direction. The calculated spectra show a flat dispersion caused by the frustration due to $J_2$ in the low energy region. On the other hand, in the high energy region, there is a dispersion, the energy position of which corresponds to that obtained from the linear spin-wave theory. The spectra of the dispersion are, however, broad. We show that this dispersion is brought about by the IC interaction with zigzag-type structure. The experimental results are discussed in the light of our theoretical results.

We first construct a minimal model for Li$_2$CuO$_2$ including IC interaction. As shown in Fig.~1(a), Li ions are located between the chains. The IC interaction works in a- and c- directions via Li ions. Since the hatched chains in Fig.~1(a) are situated in b-c plane, the orbitals relevant to the electronic states in each chain are Cu3$d_{y^2-z^2}$ and O2$p_{y,z}$ ones. The possible paths which give IC interactions along c-axis $J_c$ and along a-axis $J_a$ are shown in Fig.~1(a) by thick solid and dotted lines, respectively. For $J_c$ (solid line), the orbitals of Li ions (Li1$s$, 2$s$) couple to  O2$p_{x}$ in {\it one} chain, but to O2$p_{y,z}$ orbitals in {\it another} one. On the other hand, for $J_a$, they couple to only O2$p_{x}$ orbitals in {\it both} chains. As a result, the magnitude of $J_a$ is expected to be much smaller than that of $J_c$ due to orthogonality of Cu3$d_{y^2-z^2}$ and O2$p_{x}$ orbitals. In fact, the neutron-scattering experiment\cite{Boehm} indicates that the width of magnon dispersion along a-axis is narrower than that along c-axis, and $J_a$ is $\sim4$~K, which is less than half of $J_c$. In the present study, we neglect $J_a$ for simplicity.

\begin{figure}
\begin{center}
\epsfig{file=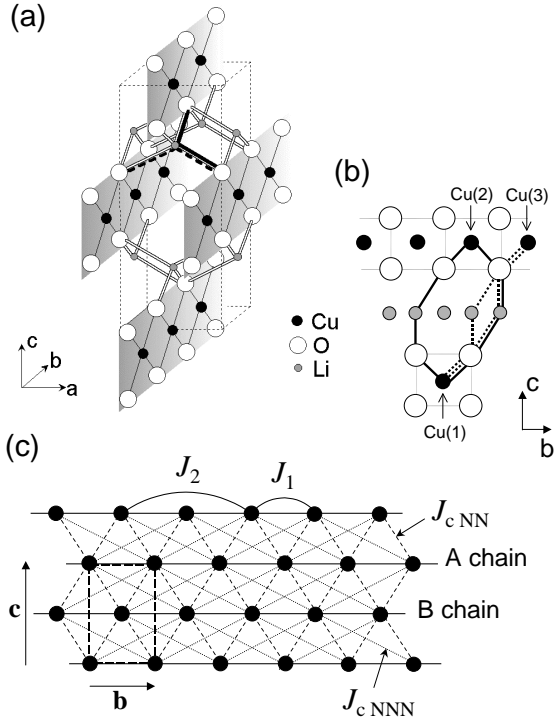,width=8.0cm,clip=}
\caption{(a) The crystal structure of Li$_2$CuO$_2$. The solid, open and hatched circles denote Cu, O and Li ions, respectively. The edge-sharing chains run along b-axis, and are stacked along a- and c-axes. The thin solid and open lines connect intra- and inter-chain neighboring ions, respectively. The thick solid and dotted lines represent paths which give the IC interactions along c- and a-axes via Li ions, respectively. (b) The schematic picture of Cu-O-Li-O-Cu paths. NN Cu ions [Cu(1) and Cu(2)] and NNN Cu ions [Cu(1) and Cu(3)] in c-direction are connected by the paths depicted by two solid and two dotted lines, respectively. (c) The $J_1$-$J_2$-$J_c$ model. The Cu sites with S=1/2 spin are depicted by the solid circles. $J_1$ and $J_2$ are interactions between NN and NNN Cu spins in chains. Two chains are coupled by IC interactions $J_{c {\rm NN}}$ and $J_{c {\rm NNN}}$, which work between NN and NNN Cu spins in different chains, respectively. We take $J_{c {\rm NN}}$=$J_{c {\rm NNN}}$$\equiv$$J_c$. ${\bf b}=b{\bf e}_b$ and ${\bf c}=c{\bf e}_c$ are the primitive vectors. The rectangle represented by the broken lines  denotes a unit cell.
}
\label{fig:1}
\end{center}
\end{figure}

For $J_c$, there are two paths connecting between NN Cu ions [Cu(1) and Cu(2)] and between NNN Cu ions [Cu(1) and Cu(3)] as shown by two solid and two dotted lines in Fig.~1(b), respectively. They give rise to zigzag-type IC interactions between NN Cu spins and between NNN ones, $J_{c {\rm NN}}$ and $J_{c {\rm NNN}}$, respectively. From the  consideration of the crystal structure and the orbital configuration, the magnitudes of $J_{c {\rm NN}}$ and $J_{c {\rm NNN}}$ are found to be the same because each path makes an equal contribution to the two interactions. Therefore, we take $J_{c {\rm NN}}=J_{c {\rm NNN}}\equiv J_c$.

\begin{figure}[t]
\begin{center}
\epsfig{file=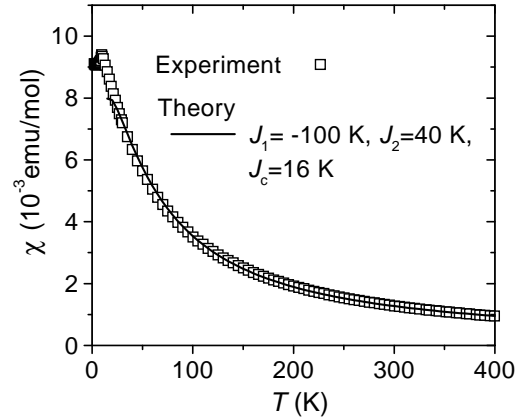,width=7.2cm,clip=}
\caption{The magnetic susceptibility $\chi(T)$ of Li$_2$CuO$_2$. The square symbols denote the experimental results in the magnetic field applied along the chains from Ref.~3. The solid curve is the theoretical result in the $J_1$-$J_2$-$J_c$ model with $J_1$=$-$100~K, $J_2$=40~K and $J_c$=16~K, obtained numerically on the system with 8$\times$2 sites. $g$-factor is taken to be 2 from Ref.~10.
}
\label{fig:2}
\end{center}
\end{figure}

Based on the above consideration, we adopt a $J_1$-$J_2$-$J_c$ model shown in Fig.~1(c). In  the previous study,\cite{Mizuno1} $J_1$ was evaluated to be $-$100~K.  We determine $J_2$ and $J_c$ based on the analysis of the experimental data of $\chi(T)$.  By  diagonalizing the Hamiltonian of the $J_1$-$J_2$-$J_c$ model for finite-size clusters with M$\times$N sites (M is the number of sites in a chain and N is the number of chains), the theoretical $\chi(T)$ is obtained. We preliminarily calculated it in 4$\times$2 and 4$\times$4 clusters in order to check the size effect along the IC direction (N dependence), and found that the difference of $\chi(T)$ between  N=2 and 4 is so small that clusters with N=2 are enough to investigate the effect of the IC interaction on $\chi(T)$. In Fig.~2, we show the result for a 8$\times$2 cluster by solid line together with the experimental data denoted by square symbols. A good agreement between theory and experiment is obtained for $J_2$=40~K and $J_c$=16~K. This value of $J_c$ is not far from the value determined by the neutron-scattering experiment.\cite{Boehm}  The theoretical results of  $\chi(T)$ reproduce well the divergent behavior of the experimental ones at low  temperatures. We examined $\chi(T)$ of the single chain (the $J_1$-$J_2$ model) in the previous study,\cite{Mizuno1} and obtained a peak in $\chi(T)$ at $T$=40~K. Therefore, we conclude that $\chi(T)$ is not explained unless $J_c$ is not taken into account.

Next, we calculate the dynamical spin correlation function $S({\bf q},\omega)$ in order to investigate the magnetic excitation. This is defined as follows.
\begin{equation}
S({\bf q},\omega)=\sum_m |\langle m|S_{\bf q}^z|0 \rangle |^2 \delta (\omega-E_m+E_0),
\end{equation}
where $|0 \rangle$ and $|m \rangle$ are the ground and excited states with energies $E_0$ and $E_m$, respectively. $S_{\bf q}^z$ is the Fourier transform of spatial spin density.

\begin{figure}[t]
\begin{center}
\epsfig{file=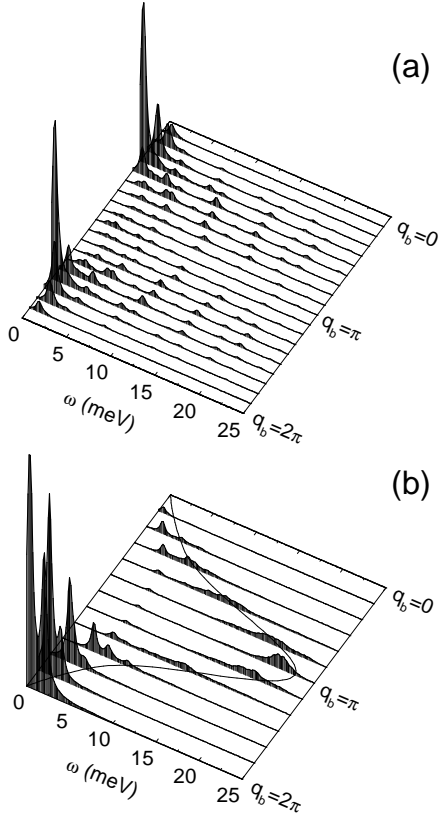,width=7.4cm,clip=}
\caption{(a) $S({\bf q}, \omega)$ along the chain direction $q_b$ with $J_1$=$-$100~K, $J_2$=40~K and $J_c$=0~K. The chain has 20 sites. (b) The same as (a) but $J_c$=16~K. The system has 12$\times$2 sites, simulating the coupled edge-sharing chains in Li$_2$CuO$_2$.  The $\delta$ functions are convoluted with a Lorentzian broadening of 0.3~meV. The thin solid curve denotes a magnon dispersion [Eq.~(3)] obtained by the linear-spin-wave theory.}
\label{fig:3}
\end{center}
\end{figure}

It is instructive to consider $S({\bf q},\omega)$ for the single chain ($J_c$=0 case) in order to understand the effect of IC interaction on the magnetic excitation. The calculation is performed by applying the exact diagonalization technique on a 20-site single chain with $J_1$=$-$100~K and $J_2$=40~K. Because the ratio of $J_2$/$|J_1|$(=0.4) is more than 1/4, the system is described as a frustrated chain. The calculated results of $S({\bf q},\omega)$ are shown in Fig.~3(a), in which $q_b$ denotes a momentum of the chain direction. The spectra have a period of $2\pi$, and are symmetric with respect to $q_b=\pi$. Since the ground state has incommensurate spin correlation, the spectral intensity is much larger at $q_b$$\sim$1/2$\pi$ and 3/2$\pi$ than that at other momenta. Much of the spectral weight is concentrated in the very low energy region ($\omega\alt2$ meV), and small weight spreads over the high energy region.

We turn our attention to the system with finite $J_c$. Since we are interested in the effect of the IC interaction on the spectra of frustrated chain, $S({\bf q},\omega)$ along the chain direction is examined. $S({\bf q},\omega)$'s for 12$\times$2 and 6$\times$4 clusters, where ${\bf q}$ is the two-dimensional vector with ${\bf q}=(q_b, q_c)$, are calculated to understand the effect of the IC interaction. By comparing results between the two clusters, we found that there is no essential difference along the chain direction. Therefore, we consider $S({\bf q},\omega)$ for the 12$\times2$ cluster.

The operator $S_{\bf q}^z$ for the 12$\times$2 cluster is written as 

\begin{eqnarray}
S_{\bf q}^z=\frac{1}{\sqrt{N_s}} \sum_i e^{iq_b y_i} (S_{A,i}^z+S_{B,i}^z e^{i(\frac{q_b}{2}+\frac{q_c}{2})}),
\end{eqnarray}
where $S_{A,i}^z$ and $S_{B,i}^z$ are the $z$ component of spin operators at $i$-th sites on $A$ and $B$ chains in Fig.~1(c), respectively. $y_i$ is the $y$ component of the  position vector at $i$-th sites, and $N_s$ is the number of Cu sites. Since in the case of finite $J_c$, the system has two Cu sites in unit cell as shown in Fig.~1(c), the lattice period in the chain direction is changed from $|{\bf b}|$ to $|{\bf b}|/2$.  Correspondingly, the period of $S({\bf q},\omega)$ is changed from $2\pi$ to $4\pi$.

The result of $S({\bf q},\omega)$ for the 12$\times$2 cluster is shown in Fig.~3(b),  where the momentum of $q_c$ is taken to be 0.\cite{note0} The spectrum is completely different from that for the single chain shown in Fig.~3(a). The strong intensity is seen at $q_b\sim2\pi$, and thus the spectrum becomes asymmetric with respect to $q_b=\pi$.
  The asymmetry is understood as follows. For example, consider the spectra at ${\bf q}=(0, 0)$ and $(2\pi, 0)$. $S_{\bf q}^z$ at these momenta is given by $\frac{1}{\sqrt{N_s}} \sum_i (S_{A,i}^z+S_{B,i}^z)$ and $\frac{1}{\sqrt{N_s}} \sum_i (S_{A,i}^z-S_{B,i}^z)$, respectively.  From these expressions, it is clear that the weight at ${\bf q}=(2\pi, 0)$ is larger than that at ${\bf q}=(0, 0)$, when the $A$ and $B$ chains are coupled antiferromagnetically. A flat dispersion with small intensity is seen around $q_b\sim\pi$ at $\omega\alt5$~meV. This is a remnant of the continuum seen in Fig.~3(a). In addition, a dispersive spectrum with energy-maximum at $q_b=\pi$ is seen in Fig.~3(b).  This dispersion is consistent with that obtained by the linear spin-wave theory in which the intrachain FM and interchain AFM ordering is assumed;
\begin{eqnarray}
\omega(q_b)&=&\sqrt{[J_1(\cos q_b -1)+J_2(\cos 2q_b -1)+8J_c]^2}  \nonumber \\
&&\overline{-64J_c^2\cos^2 q_b \cos^2 \frac{q_b}{2}}.
\end{eqnarray}
Here, we note that $J_c$ of zigzag-type is responsible for this dispersion. $J_c$ connects {\it four} spins in a chain with {\it one} spin in the neighboring chain as shown in Fig.~1(c). Therefore, these {\it four} spins tend to align parallel so as to reduce the magnetic energy. This is why the IC interactions have large contribution to the dispersion. The spectra of the dispersion are rather broad. This is clearly different from the single-peak structure in a simple FM chain. This is because FM alignment in the chain is disturbed by the quantum fluctuation caused by $J_c$ and frustration by $J_2$.

Finally, we discuss the recent inelastic neutron-scattering data along b-axis for Li$_2$CuO$_2$ in the light of our theoretical results. The experiment has shown that (i) when the momentum is far from magnetic zone center $q_b=2\pi$, the spectral intensity is very small, and (ii) the lowest excitation appears at the magnetic zone center, and the dipersion has a minimum at $q_b=\pi$.\cite{Boehm} The feature (i) is in good agreement with the momentum dependence of the spectral intensity shown in Fig.~3(b).~\cite{note1}
For the feature (ii), the structure at $\pi$ in the experiment may correspond to that at $\omega\alt5$ meV which is a remnant of the frustrated state.\cite{note2} On the other hand, the dispersion at $\omega\alt20$ meV has not been observed experimentally.  In order to find the dispersion, it will be necessary to do more detail experiment, especially, in the higher energy region.

We have studied the magnetic excitation spectra of Li$_2$CuO$_2$, and found that the zigzag-type IC interaction brings about the dramatic difference seen in the spectra between Fig.~3(a) and (b). Therefore, it is meaningful to examine the spectra of other edge-sharing compounds such as La$_6$Ca$_8$Cu$_{24}$O$_{41}$ and Ca$_{2}$Y$_2$Cu$_5$O$_{10}$ because they have different IC interactions, reflecting the difference of the crystal structures.  For example, in La$_6$Ca$_8$Cu$_{24}$O$_{41}$, which has the structure similar to Sr$_{14}$Cu$_{24}$O$_{41}$, IC interactions are expected to be smaller than that in Li$_2$CuO$_2$.\cite{Regnault,Matsuda} We thus suppose that the weight of the dispersion at high energy region ($\alt$20~meV) is very small and much of  the weight is concentrated in the low-energy region ($\alt$5~meV) in La$_6$Ca$_8$Cu$_{24}$O$_{41}$ in contrast to Li$_2$CuO$_2$.
A comparison among these materials will yield a clearer understanding of the effect of IC interactions on the magnetic excitation. 

In summary, we have investigated the magnetic properties of Li$_2$CuO$_2$ with edge-sharing chains, taking into account the interchain interaction. We determined the magnetic interactions in a chain and between chains by the analysis of the magnetic susceptibility. We also calculated the dynamical spin correlation function. The results show a dispersion with broad spectra induced by the interchain interaction. The zigzag-type interchain interaction causes the dramatic effects on the magnetic properties of Li$_2$CuO$_2$. It is highly desirable that the inelastic neutron-scattering experiment is performed in the wide energy region to understand the characteristics of the dispersion along the chain.

We would like to thank M. Matsuda, H. Eisaki and K. Mochizuki for valuable discussions. This work was supported by a Grant-in-Aid for Scientific Research on Priority Areas from the Ministry of Education, Science, Sports and Culture of Japan. The parts of the numerical calculation were performed in the Supercomputer Center, Institute for Solid State Physics, University of Tokyo, and the supercomputing facilities in Institute for Materials Research, Tohoku University.  Y. M. acknowledges the financial support of Research Fellowships of the Japan Society for the Promotion of Science for Young Scientists.


\begin{references}

\bibitem{Sapina} F. Sapi\~na, J. Rodr\'iguez-Carvajal, M. J. Sanchis, R. Ib\'a\~nez, A. Beltr\'an, and D. Beltr\'an, Solid State Commun. {\bf 74}, 779 (1990).
\bibitem{Geertsma} W. Geertsma and D. Khomskii, Phys. Rev. B {\bf 54}, 3011 (1996).
\bibitem{Mizuno1} Y. Mizuno, T. Tohyama, S. Maekawa, T. Osafune, N. Motoyama, H. Eisaki and S. Uchida, Phys. Rev. B {\bf 57}, 5326 (1998).
\bibitem{J1J2} T. Tonegawa and I. Harada, J. Phys. Soc. Jpn. {\bf 58}, 2902 (1989); R. Bursill, G. A. Gehring, D. J. J. Farnell, J. B. Parkinson, T. Xiang, and C. Zeng, J. Phys. Condens. Matter {\bf 7}, 8605 (1995). 
\bibitem{Boehm} M. Boehm, S. Coad, B. Rossli, A. Zheludev, M. Zolliker, P. B$\ddot{\rm o}$ni, D. McK. Paul, H. Eisaki, N. Motoyama, and S. Uchida, Europhys. Lett. {\bf 43}, 77 (1998).
\bibitem{Weht} R. Weht, and W. E. Pickett, Phys. Rev. Lett. {\bf 81}, 2502 (1998).
\bibitem{Nishi} M. Nishi, O. Fujia, and J. Akimitsu, Phys. Rev. B {\bf 50}, 6508 (1994).
\bibitem{Regnault} L. P. Regnault, J. P. Boucher, H. Moudden, J. E. Lorenzo, A. Hiess, U. Ammerahl, G. Dhalenne, and A. Revcolevschi, Phys. Rev. B {\bf 59}, 1055 (1999).
\bibitem{Matsuda} M. Matsuda, T. Yoshihama, K. Kakurai, and G. Shirane, Phys. Rev. B {\bf 59}, 1060 (1999).
\bibitem{Ohta} H. Ohta, N. Yamauchi, T. Nanba, M. Motokawa, S. Kawamata, and K. Okuda, J. Phys. Soc. Jpn. {\bf 62}, 785 (1993).
\bibitem{note0} In the 12$\times$2 cluster, $q_c$=0 and 2$\pi$ are defined.  Since $S({\bf q},\omega)$ at ${\bf q}=(q_b, 0)$ is equivalent to that at ${\bf q}=(2\pi-q_b, \pi)$, we show only the results with $q_c=0$.

\bibitem{note1} The experiment shows a gap due to anisotropy of magnetic interactions.\cite{Boehm}  We have checked the effect of the anisotropy on $S({\bf q}, \omega)$, and confirmed that the feature of the spectral structure is unchanged except opening of small gap.
\bibitem{note2} Because the minimum exists at $q_b=\pi$, it has been concluded that $J_1$ is AFM and $J_2$ is FM.\cite{Boehm} We also calculated $S({\bf q}, \omega)$ for the case that $J_1$ is AFM and $J_2$ is FM. In this case, the intensity at $q_b=\pi$ becomes much larger than those at other momenta, which suggests that the spectrum at $\pi$ should be observed distinctly.  However, the experimental data show rather large error bars for the dispersion along the chain.\cite{Boehm}

\end{references}
\end{document}